\documentclass{article}
\usepackage{amssymb}
\usepackage{graphicx}
\begin{document}

\title{Contraints on unified models for\\dark matter and dark energy using $H(z)$}

\author{J\'ulio C. Fabris$^{a,}$\footnote{E-mail: fabris@pq.cnpq.br}\,\,, Paulo L.C. de Oliveira$^{a,}$\footnote{E-mail: paulo\_lco@yahoo.com.br}\, and Hermano Velten$^{a,b,}$\footnote{E-mail: velten@physik.un-bielefeld.de}\\
\\
a Departamento de F\'{\i}sica,Universidade Federal do Esp\'{\i}rito Santo, 29075-910, Vit\'oria, Brazil\\
b Fakult\"at f\"ur Physik, Universit\"at Bielefeld, Postfach 100131, 33501 Bielefeld, Germany}

\maketitle
\begin{abstract}
The differential age data of astrophysical objects that have evolved passivelly during the history of the universe (e.g. red galaxies) allows to test
theoretical cosmological models through the predicted Hubble function expressed in terms of the redshift $z$, $H(z)$. We use the observational data
for $H(z)$ to test unified scenarios for dark matter and dark energy. Specifically, we focus our analysis on the Generalized Chaplygin Gas (GCG) and the viscous
fluid (VF) models. For the GCG model, it is shown that the unified scenario for dark energy and dark matter requires some priors. For the VF model
we obtain estimations for the free parameters that may be compared with further analysis mainly at perturbative level. 
\end{abstract}

\leftline{Pacs: 98.80.Cq, 98.80.-k, 98.80.Bp}

\section{Introduction}

The present cosmological observational estimations indicate that $95\%$ of the matter in the universe is composed of exotic components which do not
emit any kind of electromagnetic radiation in contrast to the ordinary forms of matter known in the laboratory. For this reason, these exotic components have been denoted the dark components of the universe. Ordinarily, these dark components are divided into two different ones: dark matter,
exhibiting zero effective pressure, and directly connected with the formation of structure in the universe; dark energy, exhibiting negative pressure,
responsible for the current stage of accelerated expansion of the universe. There are many candidates to represent dark matter (axions, neutralinos, etc.) and dark energy (cosmological constant, quintessence, etc.), all of them with important advantages and disadvantages. Since both components are detected only indirectly, through their gravitational effects, it is not excluded that the gravitational phenomena that are interpreted as manifestation of dark matter
and dark energy could also be explained by a deviation of the gravitational theory, i.e., general relativity. However, such proposal faces many theoretical and conceptual problems (conflict between local and global properties, etc.). In fact, all attempts to explain dark matter and dark energy may achieve remarkable success but, to our knowledge, all of them must also live with important drawbacks. For recent reviews, see references \cite{li-li,caldwell,bertone,padma,martin}.

In the last decade an interesting proposal to describe dark matter and dark energy has arised, the so-called unified model, where dark matter and dark energy are different manifestations of a unique fluid, keeping at same time general relativity framework untouched. The prototype of these unified models for dark energy and dark matter is the Chaplygin gas \cite{moschella}, which has been generalized in many different ways \cite{berto1,neven,fabris,MCG,GRB}. $K$-essence models\cite{mukah} , which are based on modified expression for the scalar field kinetic term, can also unify the dark sector, and the Chaplygin gas is in fact a kind of $K$-essence model. More recently, some special types of self-interacting scalar fields have been used to represent the unified scenario \cite{scalar1,scalar2}. 

Dissipative fluid using the bulk viscous pressure \cite{winfriedviscous} has also been considered as a possibility of implementing a unified model for the dark sector \cite{viscous}, leading to very interesting results \cite{barrow,santiago}. In the homogeneous and isotropic universe background a one-component viscous fluid shares the same dynamics as a GCG \cite{Szydlowski}. 

The unified models for the dark sector have been tested in many different ways, and the conclusion is that they can survive the observational
tests, but in some cases specific priors must be introduced in order to avoid tensions among different observational tests. For example, confronting
SNIa and LSS (Large Scale Structures) data, for the Generalized Chaplygin Gas (GCG) model, it seems that the unified model must be imposed from the beginning and the  parameter $\alpha$ must be restricted to positive values, in order to alleviate the tensions in the parameter estimations. In particular, $\alpha$ must be restricted to positive values, in order to have a non imaginary sound velocity \cite{hermano}, a crucial restriction when observational
tests based on perturbative analysis are included.
\par
In this paper we will consider the $H(z)$ observational data to investigate unified cosmologies using the fluid description for the cosmic components. Our candidates for the unified dark sector will be i) the Generalized Chaplygin gas, with equation of state 
\begin{equation}
p_c=-\frac{A}{\rho^{\alpha}_c},
\label{chaplyginEoS}
\end{equation}
where $A$ and $\alpha$ are constants and $\rho_c$ is its density; ii) the bulk viscous fluid that is described by the Eckart's formula
\begin{equation}
p_ v=-\Theta \xi= -\Theta \xi_0 \rho^{\nu}_v,
\label{viscousEoS}
\end{equation}
where $\Theta=3H$ is the fluid expansion and $\xi=\xi(\rho_v)=\xi_0 \rho^{\nu}_v$ is the coefficient of bulk viscosity, written in terms of the density $\rho_v$ with constant parameters $\xi_0$ and $\nu$.
\par
We will not consider, initially, any other observational test, and the reason to concentrate on only one test is to verify to what extend there is potential
tensions in the initial proposal of the model (for example, no need of an extra dark matter component). It is important, in order to have future
reliable parameter estimations using all observational tests, to let a given observational test "to speak freely", without further constraints. Later, as an application of our analysis, we will
make a joint analysis using SN Ia, for the GCG model, for the best model
obtained from the $H(z)$ analysis.
We use only one important restriction: the spatial section will be flat. The reason for this is twofold: the predictions for a flat universe are quite robust, and this seems to be corroborated by the position of the first accoustic peak in the spectrum of the anisotropy of
the cosmic microwave background radiation \cite{komatsu}; imposing a flat spatial section, we restrict the space of free parameters, avoiding unnecessary degeneracy
as consequence of many free parameters. Our result at the end is very clear: the choice of the priors affects the final parameter estimation of any Chaplygin-based cosmology. In particular, concerning the unification program different priors over the parameter $\alpha$ can lead us to distinct conclusions. For the viscous model, we obtain estimations for the free parameters of the model, leading to results that are in qualitative agreement
with the estimations obtained for the Chaplygin gas, but the unification scenario must be imposed from the beginning.
\par
The paper is organized as follows. In next section, we describe the models and the $H(z)$ observational data. In section III, the GCG model and the viscous model are confronted against the $H(z)$ data. In section IV, we present the final conclusions.

\section{The background dynamics of the unified models}

The GCG model is characterized by the equation of state (\ref{chaplyginEoS}). The GCG is a realization of the energy-momentum tensor that appears in the right
hand side of the Einsteins equations. The original interest for the Chaplygin gas came, from the theoretical side, to its connection to
string theory \cite{jackiw} and, both from the theoretical and observational point of view, from the fact that it can unify dark matter and dark energy, representing both components at same
time. Hence, it could seem unreasonable to consider the Chaplygin and an extra pressureless component. But, even in a unified scenario baryons must
be present and, moreover, we would like to verify the priors and hypothesis that must be implemented in the model in order to have consistent predictions.

If we add a matter component, which accounts for the baryons and an extra amount of pressureless dark matter, the flat
Friedmann's equation and the conservation law equations read,
\begin{equation}\label{Chabackground1}
H^2 = \biggr(\frac{\dot a}{a}\biggl)^2 = \frac{8\pi G}{3}(\rho_b+\rho_{dm} + \rho_c),
\end{equation}
\begin{equation}\label{ChaCont}
\dot\rho_b + 3\frac{\dot a}{a}\rho_b=0 \quad,\quad\dot\rho_{dm} + 3\frac{\dot a}{a}\rho_{dm}=0 \quad,\quad \dot\rho_c + 3\frac{\dot a}{a}\biggr(\rho_c - \frac{A}{\rho^\alpha_{c}}\biggl) = 0.
\end{equation}
The subscripts $b$, $dm$ and $c$ denote pressureless baryonic matter, pressureless dark matter and the Chaplygin gas, respectively. In this sense, we begin with a three fluid model, where the total pressureless component is given by $\rho_m=\rho_b+\rho_{dm}$.
\par
The conservation equations can be integrated leading to 
\begin{equation}
\rho_m = \frac{\rho_{m0}}{a^3},\quad \rho_c = \rho_{c0}\biggr(A + \frac{B}{a^{3(1 + \alpha)}}\biggl)^\frac{1}{1 + \alpha},
\end{equation}
where $B$ is an integration constant. If we fix the scales such that $a(t=t_0) = 1$, where $t_0$ is the present time, {\bf $\rho_{m0}=\rho_{b0}+\rho_{dm0}$} is the
present density for pressureless matter. Also, the GCG density can be re-written as 
\begin{equation}
\rho_c = \rho_{c0}\biggr[\bar A + \frac{1 - \bar A}{a^{3(1 + \alpha)}}\biggl]^\frac{1}{1 + \alpha},
\end{equation}
where $\rho_{c0}$ is the GCG density today, and $\bar A = A/\rho_{c0}^{1 + \alpha}$ is a parameter connected with the sound velocity of the fluid.
\par
The dimensionless density parameters are given by 
\begin{equation}
\Omega_{m0} = \frac{8\pi G}{3H_0^2}\rho_{m0}, \quad \Omega_{c0} = \frac{8\pi G}{3H_0^2}\rho_{c0},
\end{equation}
where $H_0$ is the Hubble parameter today.
Hence, the Friedmann's equation may be written as,
\begin{equation}
\frac{H^2}{H_0^2} = \frac{\Omega_{m0}}{a^3} + \Omega_{c0}\biggr[\bar A + \frac{1 - \bar A}{a^{3(1 + \alpha)}}\biggl]^\frac{1}{1 + \alpha}.
\end{equation}
Expressing now the scale factor in terms of the redshift $z$, $a = 1/(1 + z)$, we find finally the expression for the function $H(z)$:
\begin{equation}
\label{h(z)}
H(z) = H_0\sqrt{(\Omega_{b0}+\Omega_{dm0})(1 + z)^3 + (1 - \Omega_{b0}-\Omega_{dm0})\biggr[\bar A + (1 - \bar A)(1 + z)^{3(1 + \alpha)}\biggl]^\frac{1}{1 + \alpha}},
\end{equation}
where $H_0$ can be expressed in terms of the reduced Hubble parameter $h$ via $H_0 =100\,h\, Km \,s^{-1}\,Mpc^{-1}$ and we have used that $\Omega_{b0} + \Omega_{dm0} + \Omega_{c0} = 1$. The nucleosynthesis prior $\Omega_{b0}=0.042$ will be adopted.
\par
Equation (\ref{h(z)}) indicates how the Hubble parameter varies with the redshift. This is the quantity that we will be confronted with the observational data.

In order to describe the background dynamics of the bulk viscous fluid we use the equation of state (\ref{viscousEoS}) and we re-write equations (\ref{Chabackground1}) and (\ref{ChaCont}) as
\begin{equation} \label{Visbackground1}
H^2 = \biggr(\frac{\dot a}{a}\biggl)^2 = \frac{8\pi G}{3}(\rho_b+\rho_{dm} + \rho_v)
\end{equation}
\begin{equation}\label{VisCont}
\dot\rho_b + 3\frac{\dot a}{a}\rho_b=0 \quad,\quad\dot\rho_{dm} + 3\frac{\dot a}{a}\rho_{dm}=0 \quad,\quad \dot\rho_v + 3\frac{\dot a}{a}\biggr(\rho_v -3H\xi_0 \rho_{v}^{\nu}\biggl) = 0.
\end{equation}
Remark that the continuity equation for the bulk viscous fluid depends on $H$. In the absence of pressureless matter we have $H\sim \rho^{1/2}_v$ and the dynamics given by equations (\ref{Visbackground1}-\ref{VisCont}) corresponds to the set (\ref{Chabackground1}-\ref{ChaCont}) if we use the correspondence
\begin{equation}
\nu=-(\alpha+\frac{1}{2}).
\end{equation}
Then, in a one fluid model, the GCG model and the bulk viscous fluid have the same background dynamics.

Since pressureless matter was considered in the expansion factor (\ref{Visbackground1}) the correspondence described above does not occur within the cosmology considered here. Consequently, the background dynamics of the bulk viscous fluid will be different from the GCG model. The fractional density for the bulk viscous fluid will be given by the numerical solution of the continuity equation
\begin{equation}
(1+z)\frac{d\Omega_{v}(z)}{dz}-3\Omega_{v}(z)+\tilde{\xi}\Omega_{v}^{\nu}\left[\Omega_{v}(z)+(\Omega_{b0}+\Omega_{dm0})(1+z)^{3}\right]^{1/2}=0,
\label{Omegarv}
\end{equation}
where the new parameter $\tilde{\xi}$ can be expressed in terms of $\xi_0$ via $\tilde{\xi}=\frac{24\pi G \xi_0}{H_0}\left(\frac{3H_{0}^{2}}{8\pi G}\right)^{\nu}$.

Once we have solved the above differential equation for $\Omega_{v}$ the final background dynamics is
\begin{equation}
H_{v}(z)=H_0\left[(\Omega_{b0}+\Omega_{dm0})(1+z)^{3}+\Omega_{v}(z)\right],
\end{equation}
where the relation $\Omega_{b0} + \Omega_{dm0} + \Omega_{v0} = 1$ still applies.


\section{Observational constraints using $H(z)$.}

The Hubble parameter may be written in terms of the redshift parameter $z$ as
\begin{equation}
H = - \frac{1}{1 + z}\frac{dz}{dt}.
\end{equation}
Hence, to determine the function $H(z)$ requires the knowledge of the derivative of $z$. This can be achieved by identifying some
"clock" galaxies that exhibit a uniform distribution of star population. Special attention is given to red galaxies. This method has
been proposed in \cite{diferentialage} and employed to such galaxies in references \cite{verde,stern,jimenez}. We will use the data listed in reference \cite{ma} (see also \cite{ma1}).
\par
The usual $\chi^2$ statistitics will be used, computing the quantity
\begin{equation}
\chi^2 = \sum_i \frac{(H(z_i)^t - H(z_i)^o)^2}{\sigma_i^2},
\end{equation}
where the superscripts $t$ and $o$ denote the evaluated theoretical value and its corresponding observational result, while $\sigma_i$ is the
observational error bar. From this quantity, the probability distribution function $PDF$ is constructed:
\begin{equation}
{\cal P} = {\mathcal A}e^{-\chi^2/2},
\end{equation}
where $\mathcal{A}$ is a normalization factor. 
\par
The probability depends on the free parameters of the theoretical model. For the GCG model, these free parameters are $h$, $\Omega_{dm0}$, $\alpha$ and
$\bar A$. In our subsequent analysis, we will consider three main cases: the unification scenario, where the $\Omega_{m0}$ contribution is given just
for the baryon component $\Omega_{b0}$; fixing $\Omega_{dm0} = 0.25$ (the $\Lambda$CDM prior), where the GCG can be seen mainly as a dark energy candidate; leaving the parameter $\Omega_{dm0}$ as a free parameter. In this way we can consistently test the unification scenario of the GCG model.

Choosing a prior is an important part of any Bayesian calculation. This includes to choose conveniently the interval for each free parameter. Concerning the parameters $h$, $\bar{A}$ and $\Omega_{dm0}$ the natural choices are
\begin{equation}
0\leq h\leq1, \quad 0\leq \bar{A} \leq 1, \quad 0\leq \Omega_{dm0}\leq 0.958.
\end{equation}
The background dynamics of the GCG model has a fundamental dependence on the parameter $\alpha$. In principle, $\alpha$ can assume any arbitrary positive value. From the perturbative perspective, for example, a positive sound velocity requires $\alpha>0$. Even very large values, as $\alpha>200$, have been found in perturbative analysis \cite{piattella1, hermano}. Thus, it seems that we do not need to fix an upper limit to $\alpha$. On the other hand, the dynamics that has motivated GCG cosmologies, i.e., a matter dominated-like behaviour in the past followed by an accelerated one, occurs only if $\alpha > -1$. Also, some estimations in the range $\alpha<0$ have been found using SNIa \cite{colistete}. Hence, in order to explore different prior choices on $\alpha$, we adopt the general priors $\alpha > \alpha_{mim}$, where $\alpha_{min}$ assumes four different values: -10.0, -2.0, -1.0, 0.0.

We firstly discuss the results concerning the unification scenario ($\Omega_{dm}=0$). Table 1 shows the one-dimensional estimations for each free parameter. The uncertainties are calculated at $2\sigma$. The results are not sensitive to the choice of the priors. There is a high concordance around the values $h=0.71$ and $\bar{A}=0.97$. Since the final estimation of any parameter depends only on the prior information of the remaining ones, in principle the estimated value for $\alpha$ does not depends on its own prior. This is a general result in bayesian analysis and is also valid for any other parameter. But this assertion is true only if the priors do not exclude regions of high probability for
the parameter under evaluation. In any case, the uncertainties for the parameter $\alpha$ can be different as shown in table 1. This is due to the fact that the PDF is limited at $\alpha_{min}$. Consequently, the area below this function, which is used to calculate the uncertainties, depends on the prior.

\begin{table}[h]
\begin{center}
    \setlength{\arrayrulewidth}{2\arrayrulewidth}  
\begin{tabular}{|c|c|c|c|c|}\hline
$\alpha_{min}$&$0.0$&$-1.0$&$-2.0$&$-10.0$\\ \hline
$h$&$0.74^{+0.07}_{-0.06}$&$0.71_{-0.09}^{+0.08}$&$0.71^{+0.08}_{-0.11}$&$0.71^{+0.08}_{-0.11}$\\ \hline
$\bar A$&$0.97^{+0.03}_{-0.27}$&$0.98^{+0.02}_{-0.56}$&$0.97^{+0.03}_{-0.72}$&$0.97^{+0.03}_{-0.64}$\\ \hline
$\alpha$&$0.00^{+1.94}_{-0.00}$&$-0.20^{+1.85}_{-0.80}$&$-0.20_{-1.42}^{+2.00}$&$-0.20^{+2.08 }_{-1.47}$\\\hline
\end{tabular}
\caption{One-dimensional estimations of the parameters $h$, $\bar{A}$ and $\alpha$ for the unified GCG scenario ($\Omega_{dm0}=0$).}
\end{center}
\label{htable1}
\end{table}

\begin{figure}[h!]
\begin{center}
\begin{minipage}{0.3\linewidth}
\includegraphics[width=\linewidth]{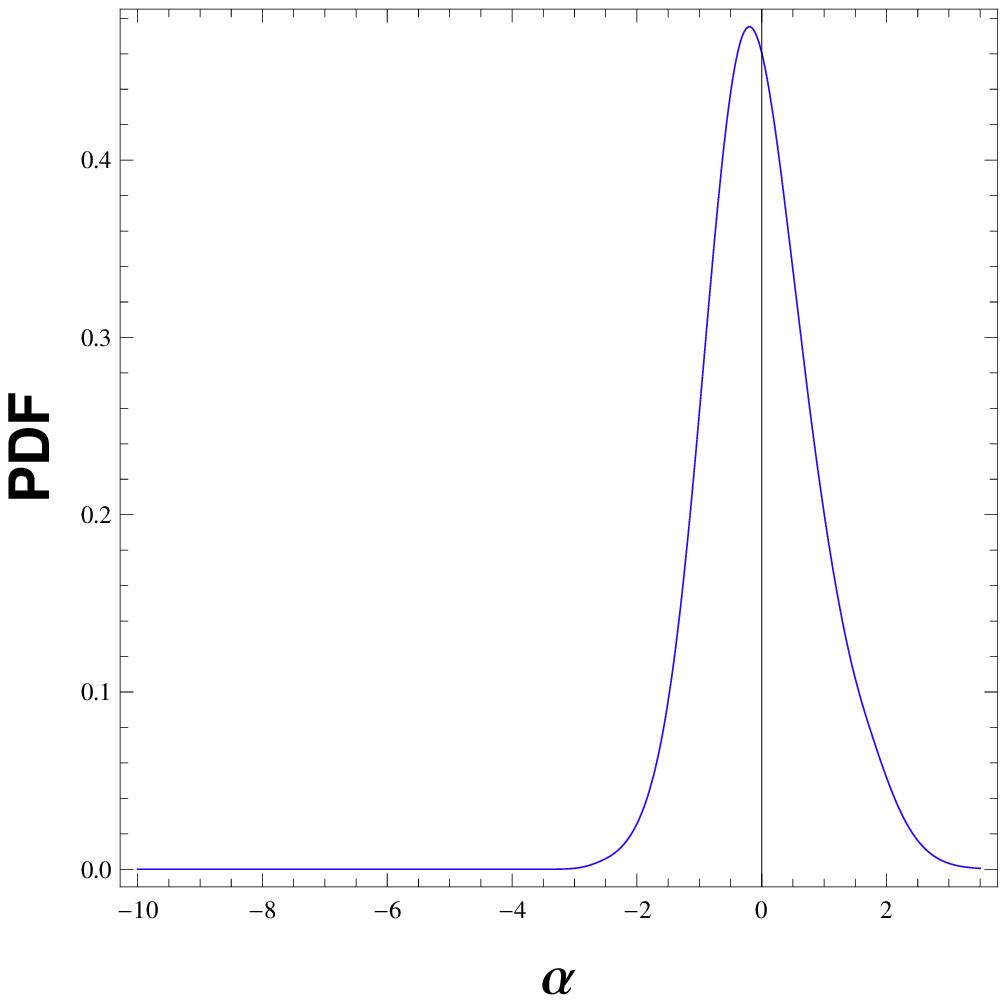}
\end{minipage} \hfill
\begin{minipage}{0.3\linewidth}
\includegraphics[width=\linewidth]{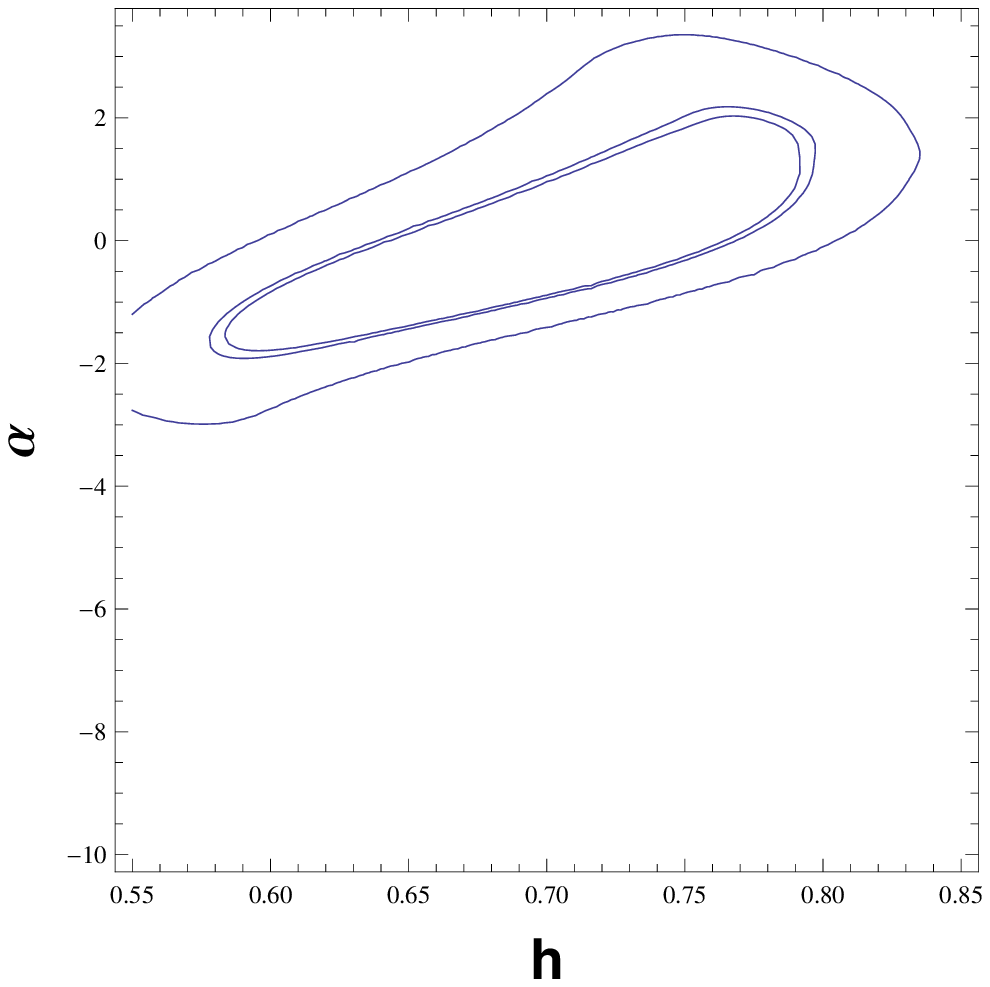}
\end{minipage} \hfill
\begin{minipage}{0.3\linewidth}
\includegraphics[width=\linewidth]{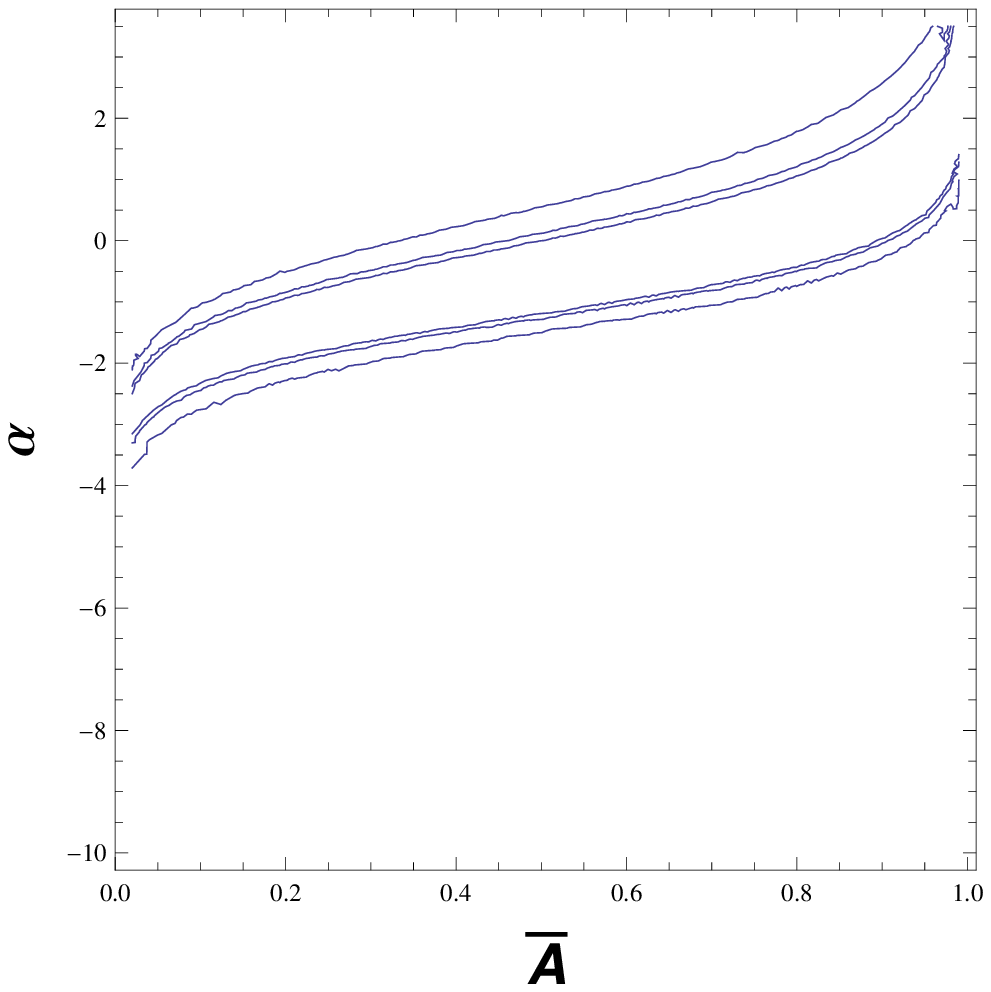}
\end{minipage} \hfill
\end{center}
\caption{{\protect\footnotesize PDFs for the unified GCG scenario ($\Omega_{dm0}=0$) if $\alpha_{min}=-10.0$.}}
\label{figure1}
\end{figure}

Figure \ref{figure1} shows a set of relevant PDFs obtained for the unified GCG model for the prior $\alpha_{min}=-10.0$. The one-dimmensional PDF for $\alpha$ (left panel) clearly shows a peak in its distribution at $\alpha=-0.20$. The contours at 1,2 and 3 $\sigma$ of confidence level in the remmaining panels are obtained after the first marginalization, i.e., we integrate over $\bar{A}$$ (h)$ in the central (right) panel. Since the contours $\alpha \times h$ (center) and $\alpha \times \bar{A}$ (left) do not depend on the $\alpha$-prior choice, we observe that a considerable probability is situated in the region $\alpha<0$. 

\begin{center}
\begin{figure}[h]
\begin{minipage}{0.3\linewidth}
\includegraphics[width=\linewidth]{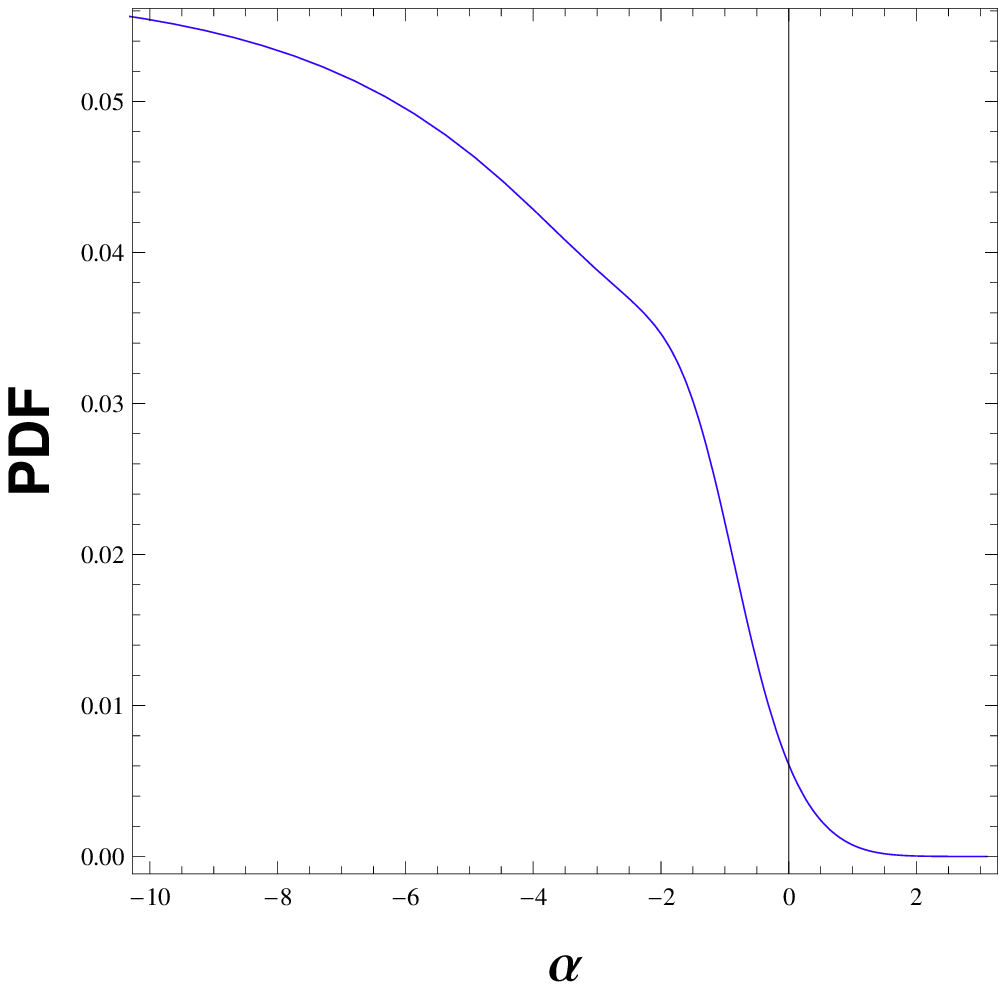}
\end{minipage} \hfill
\begin{minipage}{0.3\linewidth}
\includegraphics[width=\linewidth]{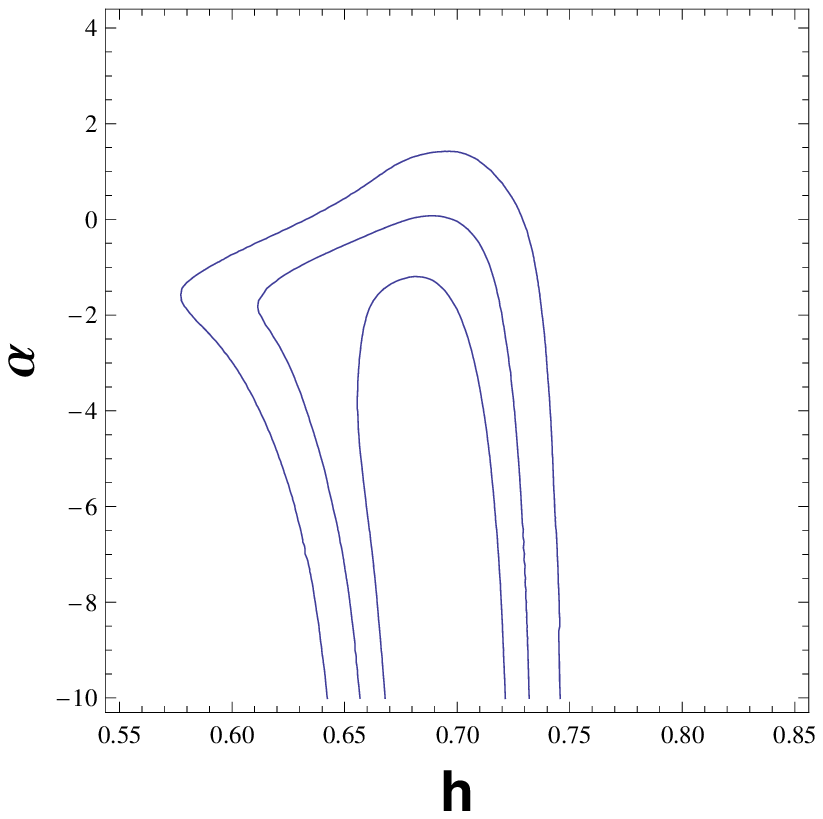}
\end{minipage} \hfill
\begin{minipage}{0.3\linewidth}
\includegraphics[width=\linewidth]{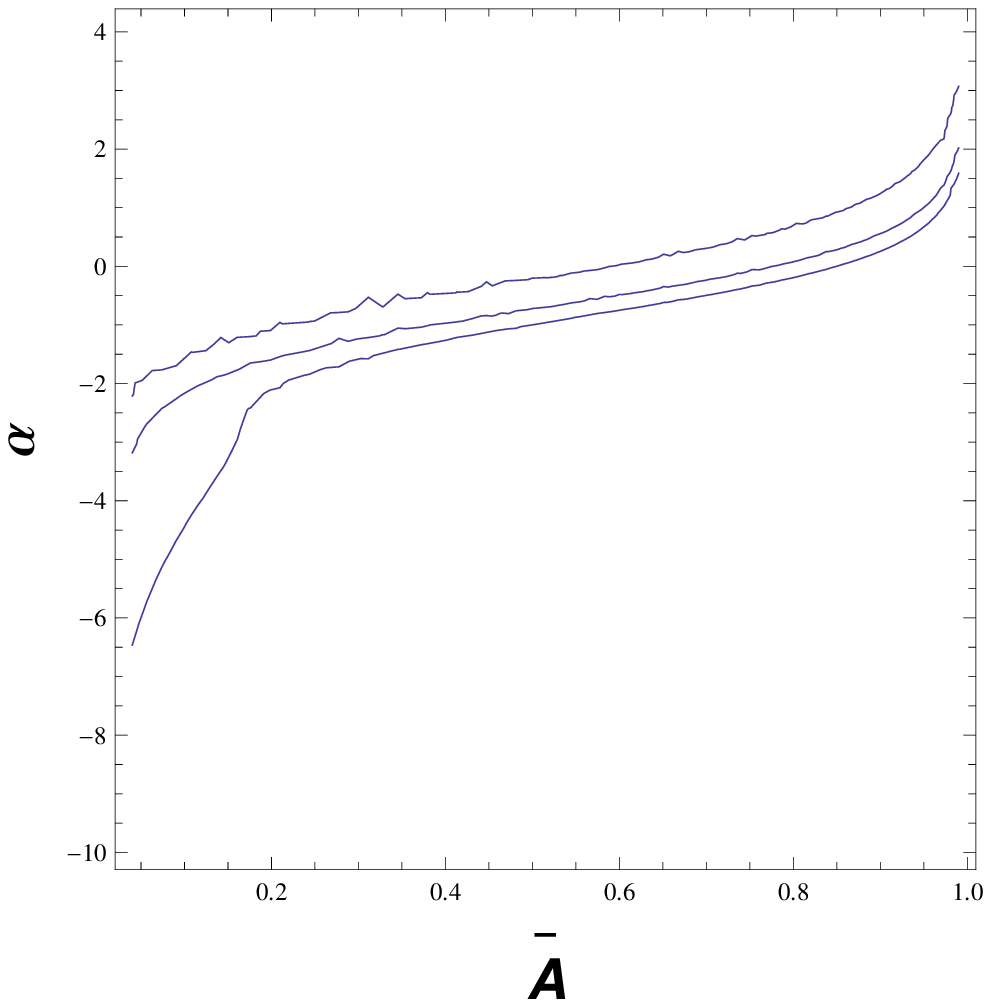}
\end{minipage} \hfill
\caption{{\protect\footnotesize PDFs for the GCG scenario with $\Omega_{dm0}=0.25$ if $\alpha_{min}=-10.0$. In the left panel we show, from botton to top, the lines are the 1, 2 and 3$\sigma$ contours of CL.}}
\label{figure2}
\end{figure}
\end{center}

In our next analysis we fix $\Omega_{dm}=0.25$ (the $\Lambda$CDM prior). The final one dimensional estimations are displayed in table 2. The influence of the $\alpha$-priors is weak and the estimations agree with $h=0.69$ and $\bar{A}=1$. However, as shown in the PDF for $\alpha$ using $\alpha_{min}=-10.0$ (left panel of figure \ref{figure2}) there does not exist a peak in the $\alpha$ distribution. Its estimated value follows the $\alpha_{min}$ value what seems to contradict its $\Lambda$CDM limit i.e. $\alpha=0$. The two-dimensional PDFs in figure 2 show explicity that large negative values for $\alpha$ are preferred if the GCG is seen as a dark energy candidate.

\begin{table}[h!]
\begin{center}
    \setlength{\arrayrulewidth}{2\arrayrulewidth}  
\begin{tabular}{|c|c|c|c|c|}\hline
$\alpha_{min}$&$0.0$&$-1.0$&$-2.0$&$-10.0$\\ \hline
$h$&$0.69^{+0.04}_{-0.05}$&$0.68_{-0.03}^{+0.04}$&$0.68^{+0.04}_{-0.08}$&$0.69_{-0.03}^{+0.07}$\\ \hline
$\bar A$&$1.00^{+0.00}_{-0.14}$&$1.00^{+0.00}_{-0.39}$&$1.00^{+0.00}_{-0.62}$&$1.00^{+0.00}_{-0.87}$\\ \hline
$\alpha$&$0.00^{+3.01}_{-0.00}$&$-1.00^{+3.25}_{-0.00}$&$-2.00^{+3.42}_{-0.0}$&$-10.00^{+6.08 }_{-0.00}$\\\hline
\end{tabular}
\caption{One-dimensional estimations of the parameters $h$, $\bar{A}$ and $\alpha$ for the GCG model with $\Omega_{dm0}=0.25$.}
\end{center}
\label{table2}
\end{table}

\begin{figure}[h!]
\begin{minipage}[h]{0.20\linewidth}
\includegraphics[width=\linewidth]{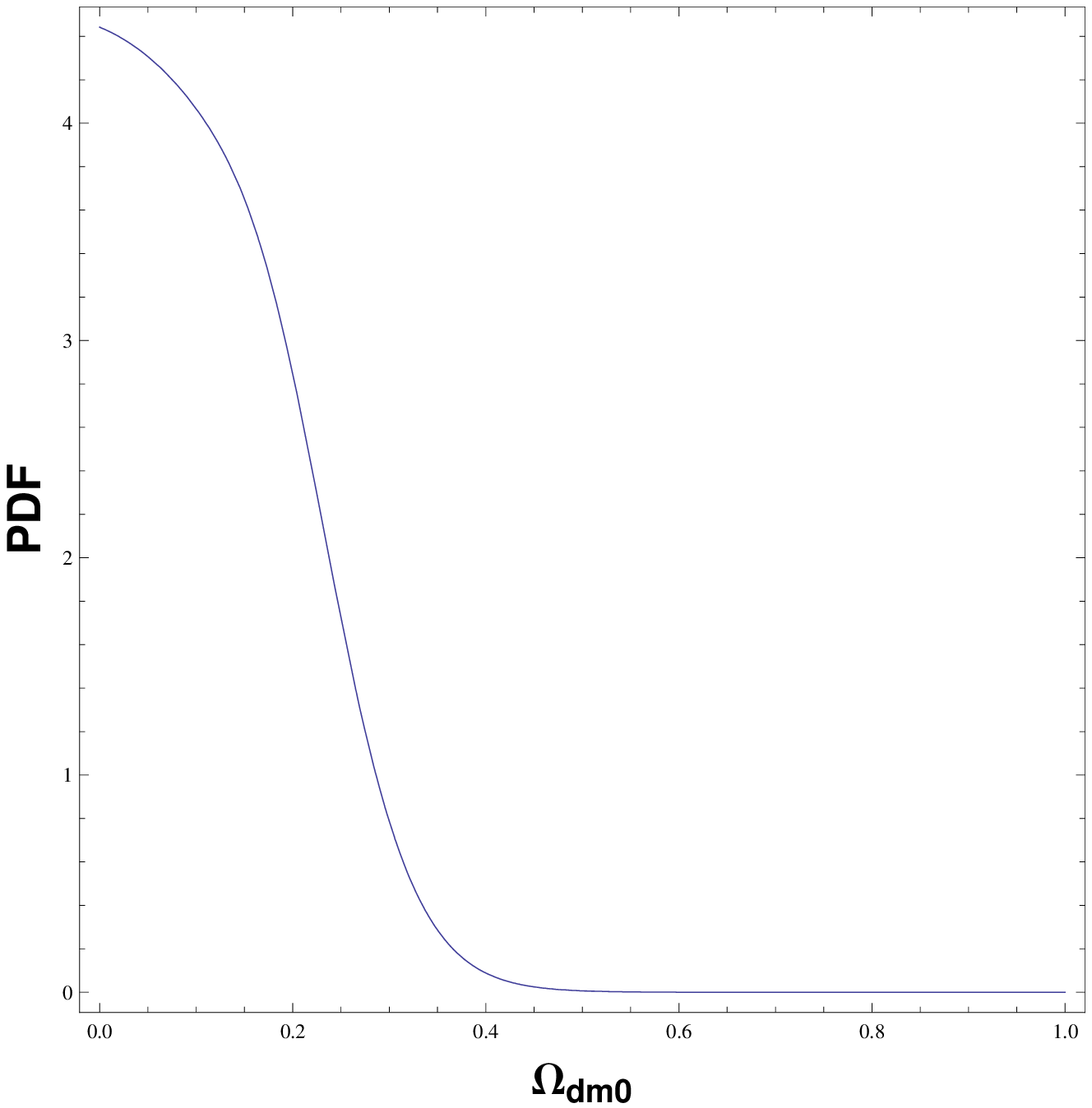}
\end{minipage} \hfill
\begin{minipage}[h]{0.20\linewidth}
\includegraphics[width=\linewidth]{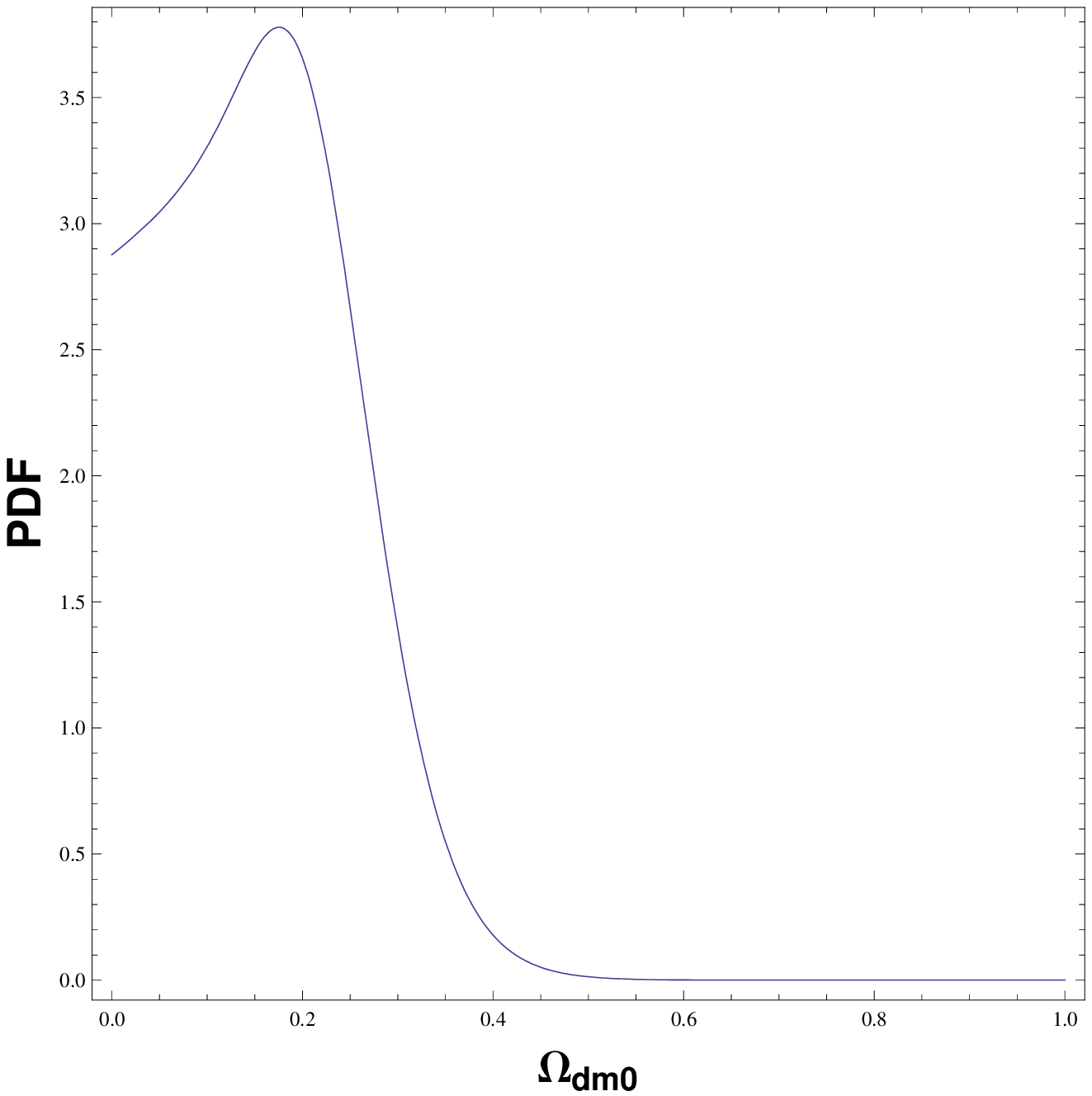}
\end{minipage} \hfill
\begin{minipage}[h]{0.20\linewidth}
\includegraphics[width=\linewidth]{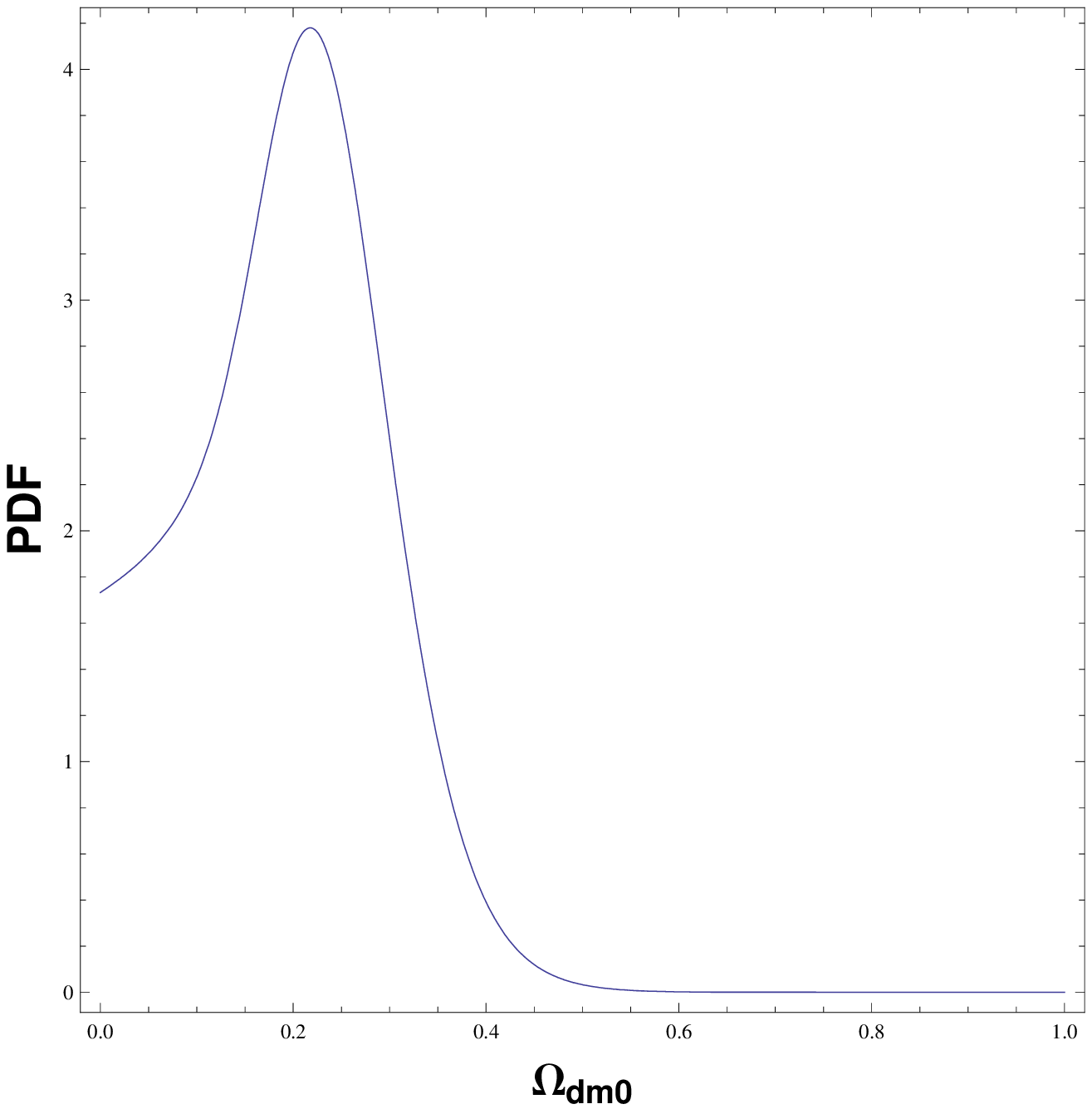}
\end{minipage} \hfill
\begin{minipage}[h]{0.20\linewidth}
\includegraphics[width=\linewidth]{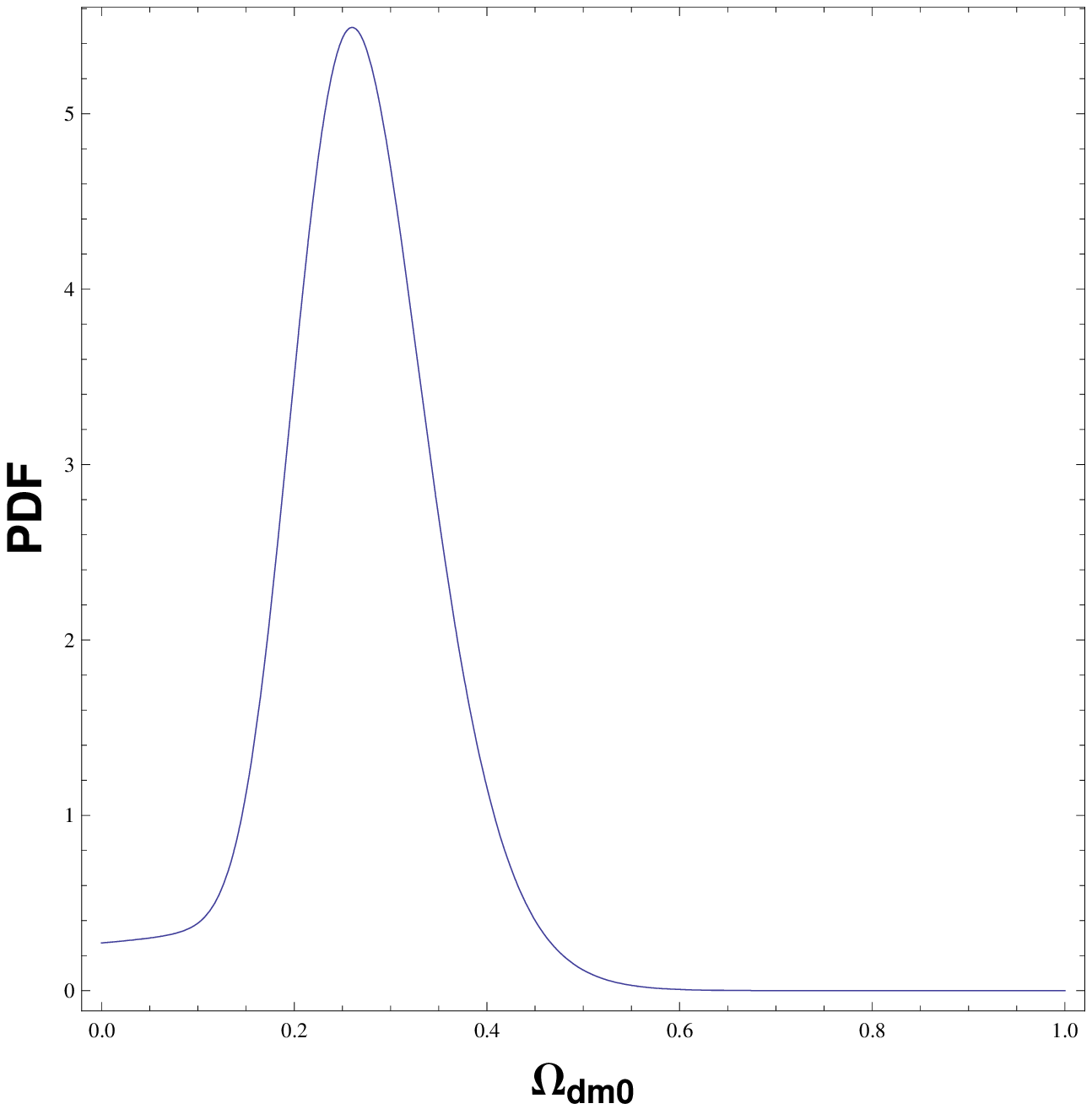}
\end{minipage} \hfill
\caption{{\protect\footnotesize One dimensional PDF for the parameter  $\Omega_{dm0}$ for different values of $\alpha_{min}$. From left to right $\alpha_{min}=0$, $\alpha_{min}=-1$, $\alpha_{min}=-2$ and $\alpha_{min}=-10$. }}
\label{Fig4}
\end{figure}

\begin{figure}[h!]
\begin{minipage}[h]{0.20\linewidth}
\includegraphics[width=\linewidth]{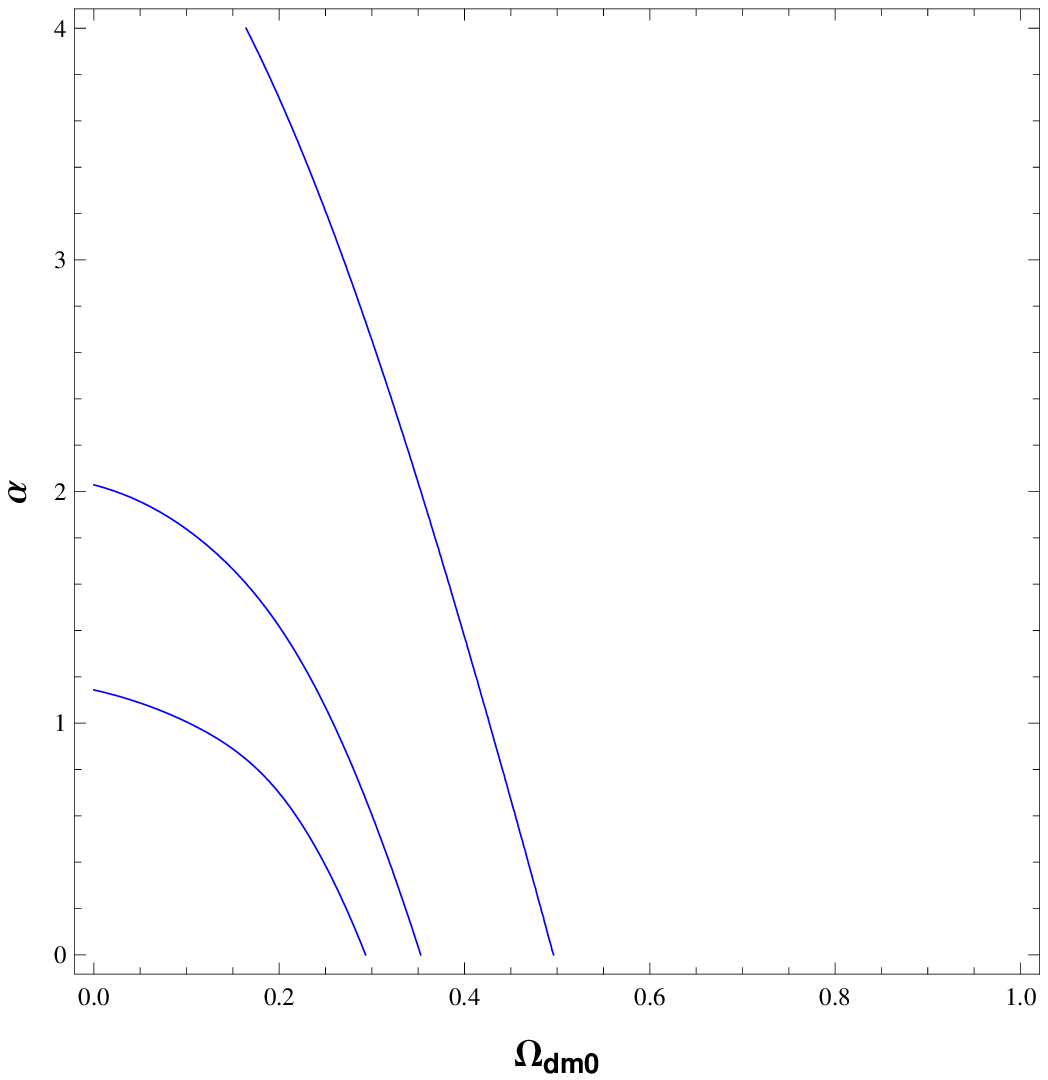}
\end{minipage} \hfill
\begin{minipage}[h]{0.20\linewidth}
\includegraphics[width=\linewidth]{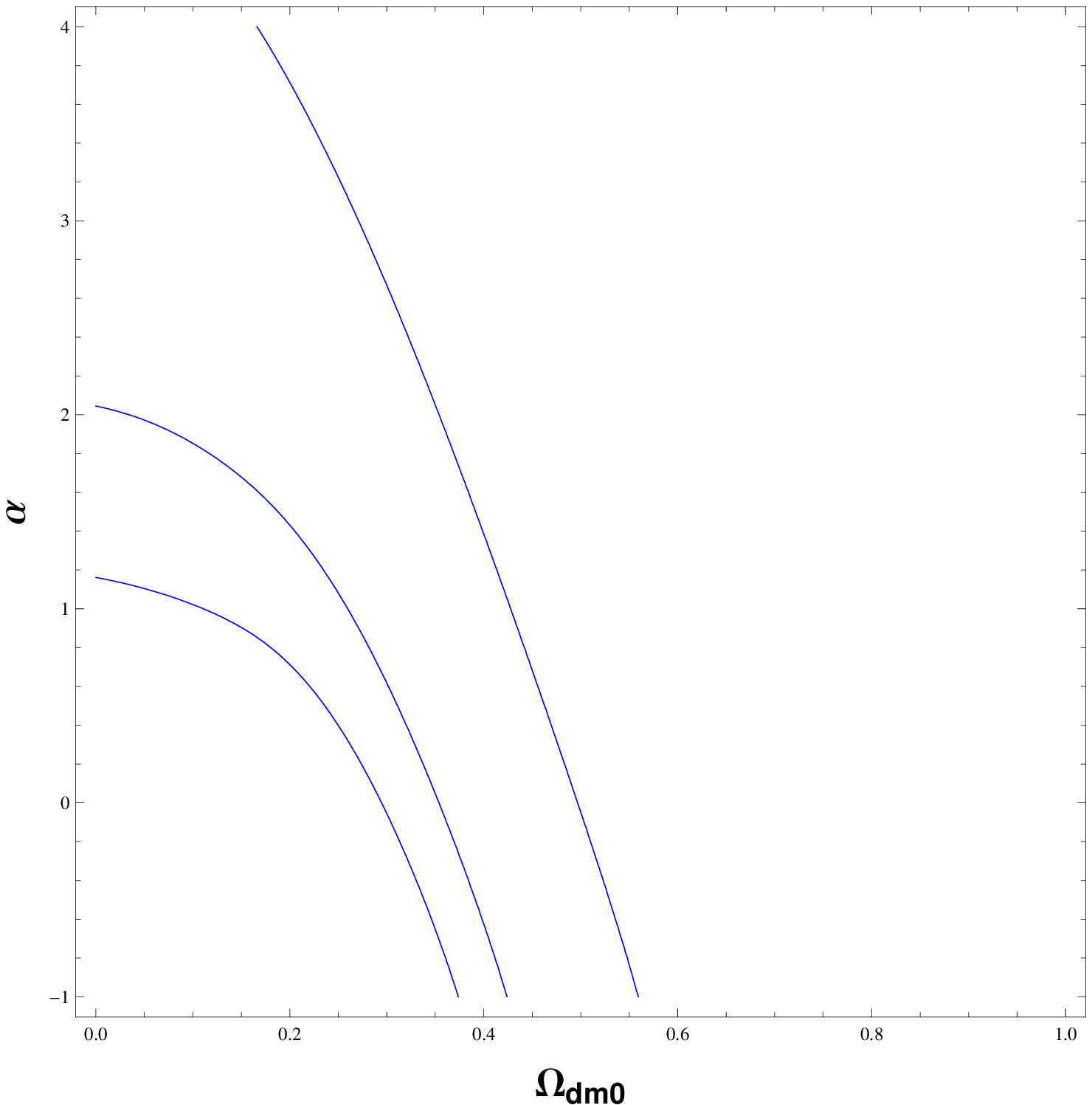}
\end{minipage} \hfill
\begin{minipage}[h]{0.20\linewidth}
\includegraphics[width=\linewidth]{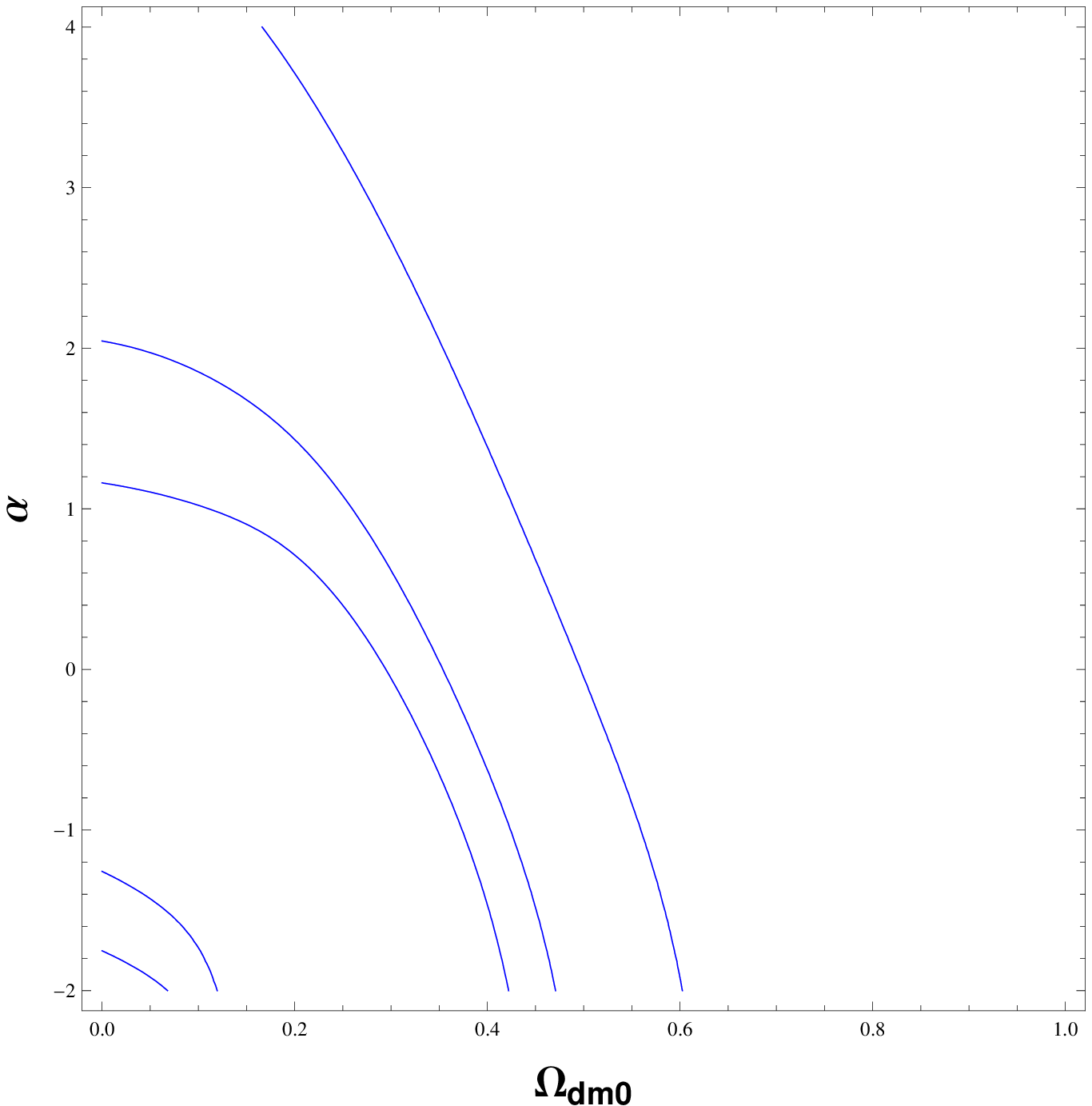}
\end{minipage} \hfill
\begin{minipage}[h]{0.20\linewidth}
\includegraphics[width=\linewidth]{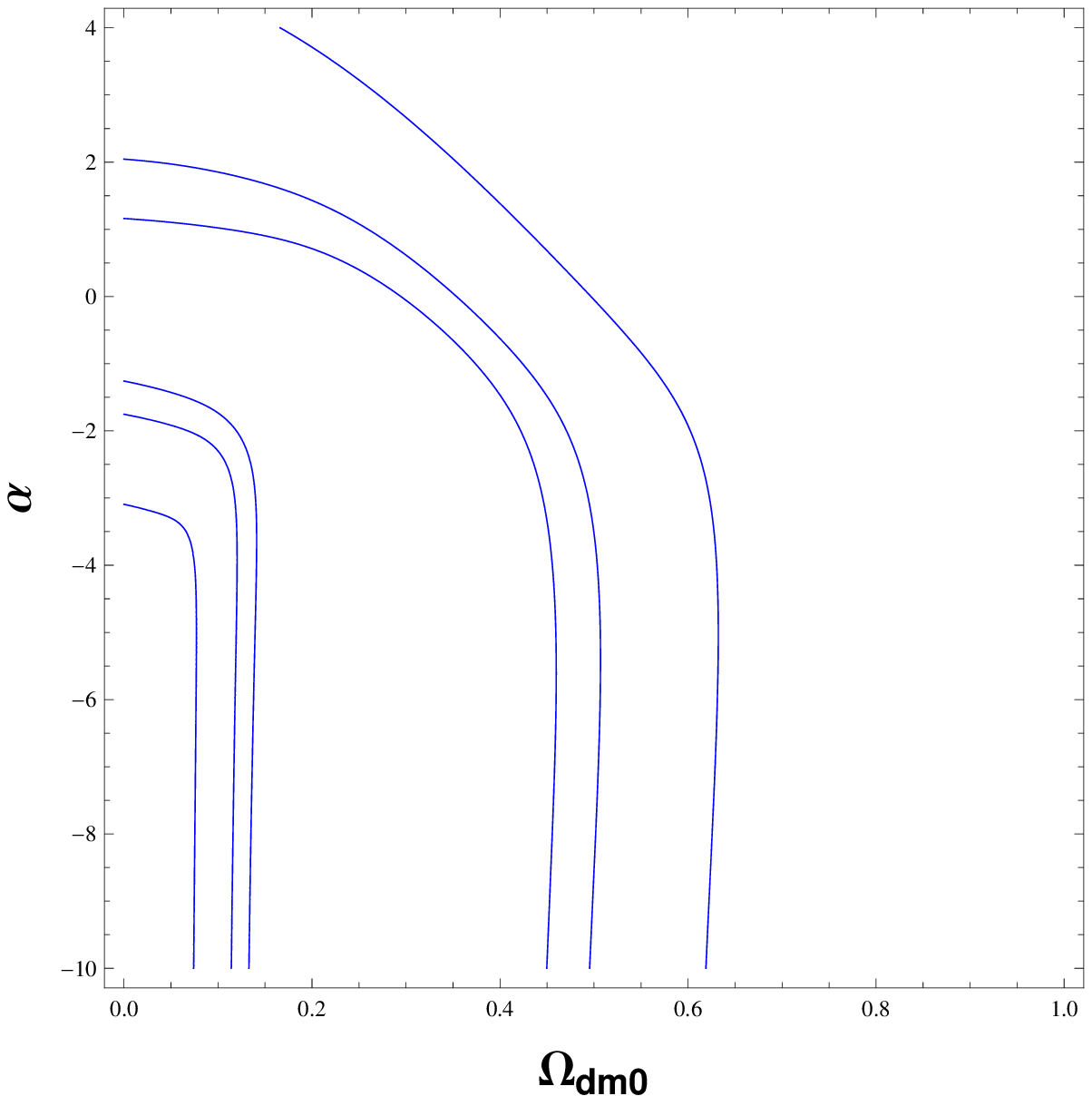}
\end{minipage} \hfill
\caption{{\protect\footnotesize Two dimensional PDF for the parameter space $\alpha \times \Omega_{dm0}$ for different values of $\alpha_{min}$. From left to right $\alpha_{min}=0$, $\alpha_{min}=-1$, $\alpha_{min}=-2$ and ($\alpha_{min}=-10$). }}
\label{Fig5}
\end{figure}

Leaving the parameter $\Omega_{dm0}$ free to vary we have a consistent way to probe whether the unification scenario is favoured ($\Omega_{dm0}=0$) or not ($\Omega_{dm0}>0$). Table \ref{table3} shows that now the choice of the prior has a fundamental role. The unification scenario is favoured only if $\alpha_{min}=0$. Figure \ref{Fig4} shows the one dimensional PDFs for $\Omega_{dm0}$ for different values of $\alpha_{min}$. The two dimensional PDFs in figure \ref{Fig5} show probability regions in the parameter space $\alpha \times \Omega_{dm0}$. When the marginalization process takes into account hypersurfaces corresponding to $\alpha\leq0$ the probability for having higher values of $\Omega_{dm0}$ increases. This is a strong evidence that the unification scenario is supported only if $\alpha_{min}=0$. 

\begin{figure}[h!]
\centering
\includegraphics[width=0.3\textwidth]{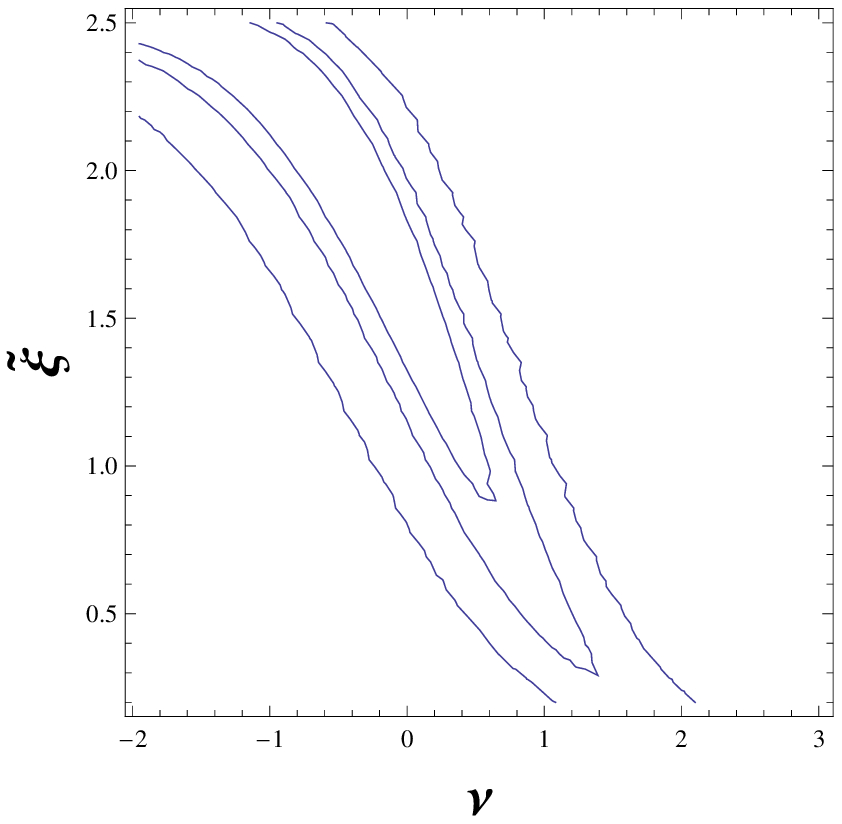}
\includegraphics[width=0.3\textwidth]{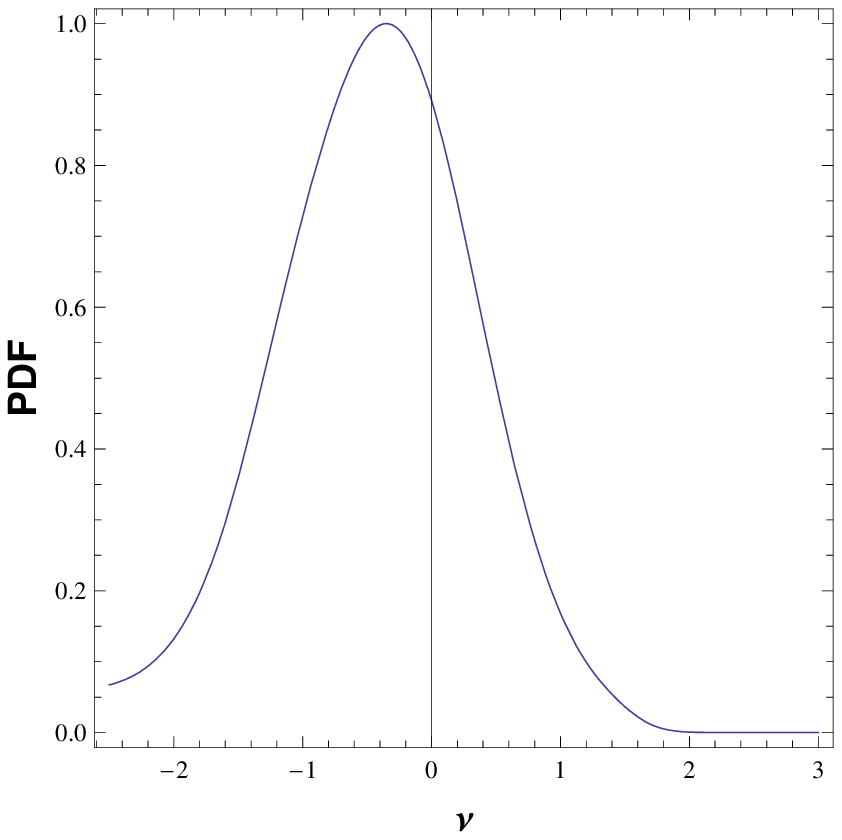}
\includegraphics[width=0.3\textwidth]{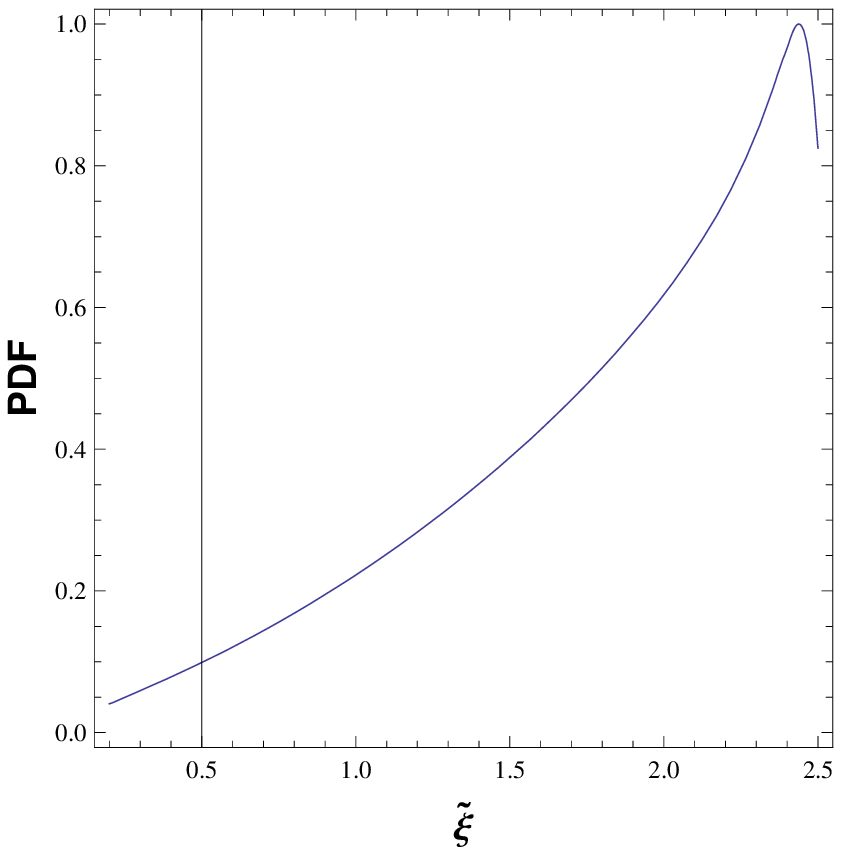}
\caption{{\protect\footnotesize PDF for the bulk viscous case.}}
\label{figviscous}
\end{figure}

Our second candidate for the unified dark sector is the bulk viscous model. For this model, we have still four free parameters when the spatial flat
condition is imposed: $\Omega_{dm0}$, $h$, $\tilde\xi$ and $\nu$. Assuming the unified condition i.e. $\Omega_{v0}=1-\Omega_{b0}$, which corresponds to the case (i) analysed above, the equation of state parameters are constrained as shown in figure \ref{figviscous}. The estimations we have obtained are: $h=0.70^{+0.06}_{-0.05}$, $\nu =-0.31^{+2.45}_{-1.62}$ and $\tilde{\xi}=2.43^{+0.17}_{-0.45}$, at $2\sigma$ CL. We remark that the only difference between this model and the GCG one is due to the existence of pressureless matter. 

The unification scenario based on a bulk viscous model is well justified when we take into account the perturbative results \cite{santiago}. In fact, concerning the GCG model the matter power spectrum indicates that the unified scenario is not the preferred one (the estimated value for the pressureless matter is $\Omega_{dm}\sim 1$) \cite{hermano}. On the other hand, the same perturbative analysis for the bulk viscous models reveals the opposite result \cite{santiago}. Hence, it seems contradictory to extend the analysis (ii) and (iii) to the bulk viscous model. Concerning the choice of the priors of the bulk viscous model there is no clear restriction from the perturbative dynamics. At first order the bulk viscous fluid behaves as a nonadiabatic component and its perturbations evolve in a very different way from the standard adiabatic scenario (which is the case of the GCG) \cite{viscous}. Then, in principle, the parameter $\nu$ can assume any real value. We remark, however, that at our knowledge no extensive quantitive parameter estimation has
been made for the viscous unified model using matter power spectrum. 

\begin{table}[h!]
   \centering   
    \setlength{\arrayrulewidth}{2\arrayrulewidth}  
      \begin{tabular}{|c|c|c|c|c|}  
      \hline
      $\alpha_{min}$ & 0.0 & -1.0 & -2.0 & -10.0 \\
      \hline
        $h$     & $0.71^{+0.07}_{-0.07}$ & $0.68^{+0.07}_{-0.08}$  & $0.68^{+0.08}_{-0.09}$  & $0.68^{+0.07}_{-0.09}$  \\ 
      \hline
      $\bar{A}$ & $1.00^{+0.00}_{-0.23}$ & $1.00^{+0.00}_{-0.49}$ & $1.00^{+0.00}_{-0.68}$ & $1.00^{+0.00}_{-0.91}$ \\
      \hline
      $\Omega_{dm0}$ & $0.00^{+0.29}_{-0.00}$ & $0.18^{+0.15}_{-0.18}$ & $0.22^{+0.14}_{-0.22}$ & $0.26^{+0.19}_{-0.15}$ \\
      \hline $\alpha$&$0.00^{+3.01}_{-0.00}$&$-1.00^{+3.25}_{-0.00}$&$-2.00^{+3.42}_{-0.00}$&$-10.00^{+6.08 }_{-0.00}$\\ 
      \hline
   \end{tabular}   
   \caption{One-dimensional estimations of the parameters $h$, $\bar{A}$ and $\Omega_{dm0}$.}
   \label{table3}
\end{table}

\begin{figure}[h!]
\centering
\includegraphics[width=0.3\textwidth]{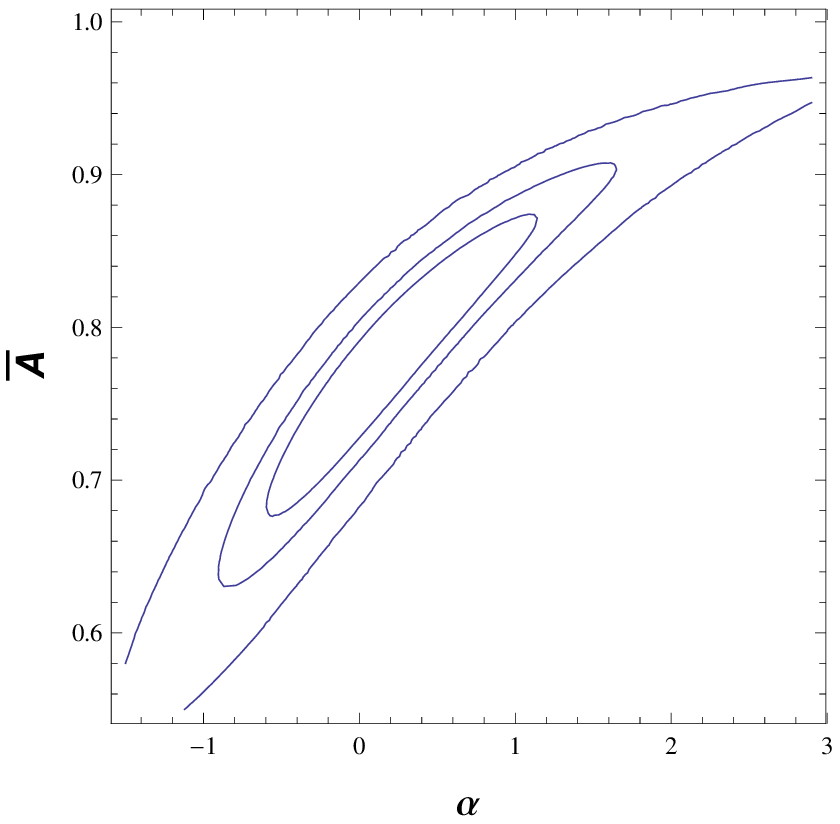}
\includegraphics[width=0.3\textwidth]{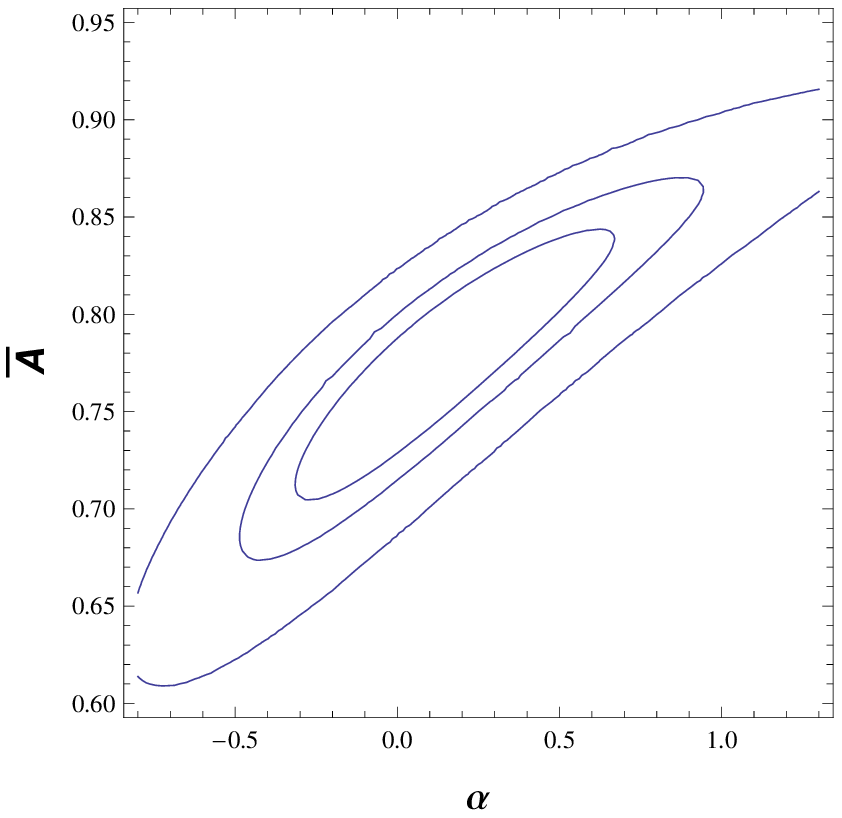}
\caption{{\protect\footnotesize PDF for the analysis using SN data and considering the unification scenario for the GCG model. Left: Using only Supernovae data. Right: The joint analysis SN+H(z). }}
\label{figjoint}
\end{figure}

From the analysis done so far using the $H(z)$ data it seems that the unification scenario with the prior $\alpha_{min}=0$ is the favoured one for Chaplygin-type cosmologies. Keeping this result in mind, we have also fixed the unification case for the bulk viscous fluid above. However, using other cosmological probes we can place stronger constraints on this unification case. In particular, a joint analysis using also the Sypernovae data will provide reliable constraints on the equation of state parameters $\alpha$ and $\bar{A}$. In order to implement the joint analysis for the Generalized Chaplygin gas we rewrite $\chi^{2}$ (equation 17) as $\chi^{2}=\chi^{2}_{SN}+\chi^{2}_{H}$, where $\chi^{2}_{H}$ corresponds to equation (16). The Supernovae data set used in our analysis is the Union2 \cite{union2}.
In figure \ref{figjoint} we assume the unification scenario ($\Omega_{dm0}=0.0$) and we show the contours for the equation of state parameters using only the Supernovae data (left panel) and the joint analysis (right panel). The contours obtained for the joint analysis are wider as the errors added to the likelihood function (compared to the increase of the total $\chi^{2}$) become now larger \footnote{The widening of the errors when the $H(z)$ and $SNIa$ data are crossed may
indicate a tension between the data even taking into account the priors used.
If there is this tension, it is small, and perhaps less important that
other controversial aspects of the SNIa data, like the calibration issues, for
example, those coming from the
use of SALT or MLCS calibration methods}. Also, note that the two samples predict a different estimated value for $\alpha$ yielding to a larger propagation of the uncertainties. For the joint analysis, we marginalize over the parameter $\bar{A}$ in order to obtain an estimation for $\alpha$ and we obtain $\alpha= 0.10^{+0.73}_{-0.57}$ which agrees (including the errors) with the result considering only the H(z) data. 

One important issue concerning parameter estimations for unified models is to discuss
the constraints from the CMB physics. The allowed parameters we have found in this work is an acceptable fit CMB data according to previous estimations \cite{finelli,schwarz}, which predict $\alpha \sim 0$. Analysis using just the position of
the first peak, also favors $\alpha \sim 0$ \cite{bertolami}. However, a complete analysis using the CMB data, in the
sense of the analysis here developped for the $H(z)$ data, presents new challenges. For example, in the references \cite{finelli,schwarz}, the parameter
$\alpha$ has been restricted to positive values and only adiabatic perturbations were considered. Since the full perturbative analysis, at least at linear level, is necessary in
order to obtain the CMB restrictions to GCG model parameters, some aspects already mentioned above point towards the necessity of a deeper analysis: the perturbative estimations for $\alpha$ show another peak of probability for very large $\alpha$, either for matter power spectrum \cite{hermano} and the Integrated Sachs Wolf effect \cite{piattella1}, with this new peak located around $\alpha \sim 200$; moreover, at perturbative level, the extension of the estimations to
negative values of $\alpha$ is problematic, due to problems concerning imaginary sound speed. Some proposals to surmount this last difficult have
been made in reference \cite{scalar1, scalar2} but no clear scenario allowing to consider negative values of $\alpha$ has emerged until now. Then even if there is an overall agreement of the estimations made here and previous estimations obtained using CMB, the restrictions on the model parameters are not the
same, and a new complete extensive analysis is required. Hence, the addition of other cosmological probes does not provide a substantial change in the final parameter estimation.   

\section{Conclusions}

The GCG model has four free parameters, when the flatness of the spatial section is imposed. Leaving the four parameters free, the unification
scenario for dark matter and dark energy is not favored: the one-dimensional probability distribution for $\alpha$ extend to arbitrary negative values (if $\alpha_{min}=-10$), and at same time the prevision for $\Omega_{m0}$ becomes similar to the $\Lambda$CDM model. In fact, the crossing of these results with the ones comming from the perturbations (the matter power spectrum) \cite{hermano}, allows to conclude: for any Chaplygin-based cosmology if the parameter $\Omega_{m0}$ is free to vary the unification scenario, with $\Omega_{dm0} \sim 0$, is not favored. However, this problem can be alleviated if the allowed range of the parameter $\alpha$ is restricted to $\alpha>0$. With this choice, the unification scenario becomes the preferred one (see figure 3). Remark that this prior is obligatory for the matter power spectrum analysis in order to have real values for the
sound speed. Again for the $H(z)$ test, fixing $\Omega_{m0} = 0.25$, we also find a plateau for negative values of $\alpha$. This result can be understood by remembering that large negative values of $\alpha$ implies a model very similar to the $\Lambda$CDM one. The imposition from the beginning of the unification prior ($\Omega_{m0} = 0.04$) leads to a more consistent scenario, with a
peak of probability for $\alpha$ near $\alpha = 0$ for any choice of $\alpha_{min}$. Note that a joint analysis using also Supernovae as cosmological probes does not change the main predictions for this case. This is consistent with the corresponding power spectrum analysis,
which implies also a peak of the probability around $\alpha = 0$. In this sense, in order to allow the parameter $\alpha$ to assume negative values the prior of the unification scenario seems to be the only consistent one.

The bulk viscous model appears as a second candidate for the unification scenario. Its background dynamics is similar to the GCG one. However, unlike the GCG, the perturbative analysis of the bulk viscous fluid supports the unification scenario. Having in mind the similarities between the bulk viscous model and the GCG one (section two), the constraints obtained in figure \ref{figviscous} agrees with the results for the GCG analysed in case (i). Due to the fact that the bulk viscous fluid has less pathologies or restrictions than the GCG one it seems put it as a more natural candidate for the unification scheme. One difficult with such model (based on the Eckart's formula (\ref{viscousEoS})) appears when dealing with the Integrated Sachs-Wolfe effect \cite{barrow}. However, the viscous pressure has an alternative description in terms of the second order corrections to the equilibrium, namely the Mueller-Israel-Stewart theory \cite{Israel}. This approach is very promissing for further investigation.

{\bf Acknowledgement} We thank CNPq (Brazil) for partial
financial support.


\end{document}